\documentclass{PoS}

\title{Mass loss of rotating stars at very low metallicity}

\ShortTitle{Mass loss at low Z}

\author{\speaker{Georges Meynet}\\
        Astronomical Observatory of Geneva University\\
        E-mail: \email{georges.meynet@obs.unige.ch}}
        
\author{André Maeder\\
        Astronomical Observatory of Geneva University\\
        E-mail: \email{Andre.Maeder@obs.unige.ch}} 
        
\author{Raphael Hirschi\\
        Dpt. of Physics and Astronomy, Basel University\\
        E-mail: \email{Raphael.Hirschi@unibas.ch}} 
        
\author{Sylvia Ekström\\
        Astronomical Observatory of Geneva University\\
        E-mail: \email{Sylvia.Ekstrom@obs.unige.ch}}        
        
\author{Cristina Chiappini\\
        Osservatorio Astronomico di Trieste and Astronomical Observatory of Geneva University\\
        E-mail: \email{Christina.Chiappini@obs.unige.ch}}                        

\abstract{Some indirect observations, as the high fraction of Be stars\footnote{stars with rotational velocities near the critical limit.} at low metallicity, or the
          necessity for massive stars to be important sources of primary nitrogen, seem to indicate
          that very metal poor stars were fast rotators. As a consequence of this fast rotation, these stars,
          contrarily to current wisdom, might lose large amounts of mass during their lifetime. 
          In this paper,
          we review various mechanisms triggered by rotation which may induce strong mass loss
          at very low metallicity. 
          The most efficient process comes from 
          surface enrichments in CNO elements which then drive mass loss by stellar winds.
          Due to this process, a fast rotating 60 M$_\odot$ with metallicities in the range of $Z=10^{-8}$
          and $10^{-5}$, can lose between 30 and 55\% of its initial mass.
          This rotationally wind ejected material participates to the chemical evolution of the
          interstellar medium, enriching it exclusively in H- and He-burning products. In particular,
          metal poor fast rotating stars may play a key role for explaining the origin of 
          the peculiar abundance pattern observed at the surface of 
          the extremely metal-poor C-rich stars, for explaining the chemical inhomogeneities observed
          in globular clusters, and the presence of stars
          in $\omega$ Cen with a very high helium content .\
          }

\FullConference{International Symposium on Nuclear Astrophysics --- Nuclei in the Cosmos --- IX\\
		 June 25-30 2006\\
		 CERN, Geneva, Switzerland}

\begin{document}

\section{Radiation driven stellar winds from non-rotating metal poor stars}

Current wisdom considers that very metal poor stars lose no or very small amounts
of mass through radiatively driven stellar winds. This comes from the fact that when the metallicity
is low, the number of absorbing lines is small and thus the coupling between the radiative forces and the matter is weak. Wind models imposes a scaling relation of the
type 
\begin{equation}
\dot M(Z)=\left({Z \over Z_\odot} \right)^\alpha\dot M(Z_\odot),
\end{equation}
where $\dot M(Z)$ is the mass loss rate when the metallicitity is equal to $Z$ and $\dot M(Z_\odot)$
is the mass loss rate for the solar metallicity, $Z$ being the mass fraction of heavy elements.
In the metallicity range from 1/30 to 3.0 times solar,
the value of $\alpha$ is between 0.5 and 0.8 according to stellar wind models (\cite{K87}; \cite{L92}; \cite{V01}). 
Such a scaling law implies for instance that 
a non-rotating 60 M$_\odot$ with $Z=0.02$
ends its stellar life with a final mass of 14.6 M$_\odot$, the same model with a metallicity of
$Z=0.00001$ ends its lifetime with a mass of 59.57 M$_\odot$ (cf. models of \cite{MMXI} and \cite{MMVIII} with $\alpha=0.5$). 

Thus
one can expect that the metal poor 60 M$_\odot$ star will give birth to a black hole.
In that case, the whole stellar mass may disappear in the remnant preventing the star from enriching the
interstellar medium in new synthesized elements. The 
metal rich model will probably leave a neutron star and contribute to the enrichment of the ISM through both
the wind and supernova ejecta. 
Let us note that a star which loses a lot of material by stellar winds may differently enrich the interstellar medium
in new elements, compared to a star which would have retained its mass all along until the supernova explosion. As \cite{M92} pointed out, when the stellar winds are strong, material partially processed
by the nuclear reactions will be released, favoring some species (as helium and carbon which would be otherwise partially destroyed if
remained locked into the star) and disfavoring other ones (as e.g. oxygen which would be produced by further transformation of the species which are wind-ejected).

Thus mass loss has very important consequences, but as said above, one expects no or very weak stellar winds
at low metallicity from non-rotating stars. Now, one knows that stars are rotating and that rotation
may change all the outputs of the stellar models, in particular the way they are losing mass.
Indirect indications, as the presence of numerous Be stars (stars near the critical limit) in metal poor clusters (\cite{MGM}), the necessity for massive stars to be efficient producers of primary nitrogen (\cite{CMB05}; \cite{CHM06}) point toward a higher proportion of fast rotators in metal poor regions.
Thus it appears worthwhile to reconsider the question of the quantity of mass lost by fast rotating stars in metal poor region.
This is the point
we want to address in this paper.

\section{General effects of rotation}

Rotation induces many processes in stellar interior (see the review by \cite{ARAA}).
In particular, it drives instabilities which transport angular momentum and chemical species.
Assuming that
the star rapidly settles into a state of shellular rotation (constant angular velocity 
at the surface of isobars), the transport equations due to meridional currents and shear instabilities
can be consistently obtained (\cite{Z92}). Since the work by J.-P.~Zahn, various improvements have been brought to the
formulas giving the velocity of the meridional currents (\cite{MZ98}), those of the various diffusive coefficients 
describing the effects of shear turbulence (\cite{M97}; \cite{TZ97}; \cite{M03}; \cite{MPZ04}), as well as the effects of rotation on the mass loss (\cite{O96}; \cite{M99}; \cite{MMVI}). 

Let us recall a few basic results obtained from rotating stellar models:

1) Angular momentum is mainly transported by the meridional currents. In the outer part
of the radiative envelope these meridional currents transport angular momentum outwards.
During the Main-Sequence phase, the core contracts and the envelope expands. The meridional currents
imposes some coupling between the two, slowing down the core and accelerating the outer layers.
In the outer layers, the velocity of these currents becomes smaller when the density gets higher, {\it i.e.},
for a given initial mass, when the metallicity is lower.

2) The chemical species are mainly transported by shear turbulence (at least in absence of a
magnetic fields; when a magnetic fields is amplified by differential rotation as in the Tayler-Spruit
dynamo mechanism \cite{S02}, the main transport mechanism is meridional circulation \cite{magn3}). 
During the Main-Sequence
this process is responsible for the nitrogen enhancements observed at the surface of OB stars (see e.g.
\cite{HL06}). The shear turbulence is stronger when the gradients of the angular velocity are stronger. Due to point 1 above, the gradients of $\Omega$ are stronger in metal poor stars and thus the mixing of the chemical
elements will be stronger in these stars. Some observations indicate that this might well be the case (\cite{V99}; \cite{VP03}). Let us note also that
the efficiency of the mixing will vary from one element to another. If an element is strongly and rapidly built up in the convective core, it will diffuse by rotational mixing more rapidly in the radiative envelope than an element with a smoother gradient between the convective core and the radiative envelope. This explains
why the stellar surface will be more rapidly enriched in nitrogen than in helium.

In addition to these internal transport processes, rotation also modifies the physical properties
of the stellar surface. Indeed the shape of the star is deformed by rotation (a fact which is now put in evidence 
observationally thanks to the interferometry, see \cite{Do03}). Rotation implies a non-uniform brightness (also now
observed, see \cite{Do05}).
The polar regions are brighter that the equatorial ones. This is a consequence of the hydrostatic
and radiative equilibrium (von Zeipel theorem \cite{Z24}). In addition, as a result of the internal transport processes,
the surface velocity and the surface chemical composition are modified.

The various processes described above are worthwhile to keep in mind since they all play some role
in promoting mass loss in metal poor stars. 
We can classify the effects of rotation on mass loss in three categories.

\begin{enumerate}
\item The structural effects of rotation.
\item The changes brought by rotation on the radiation driven stellar winds.
\item The mechanical wind induced by rotation at break-up.
\end{enumerate}

Let us now consider in turn these various processes.

\section{Structural effects of rotation on mass loss}

\begin{figure}[tb]
\begin{center}
  \resizebox{10cm}{!}{\includegraphics{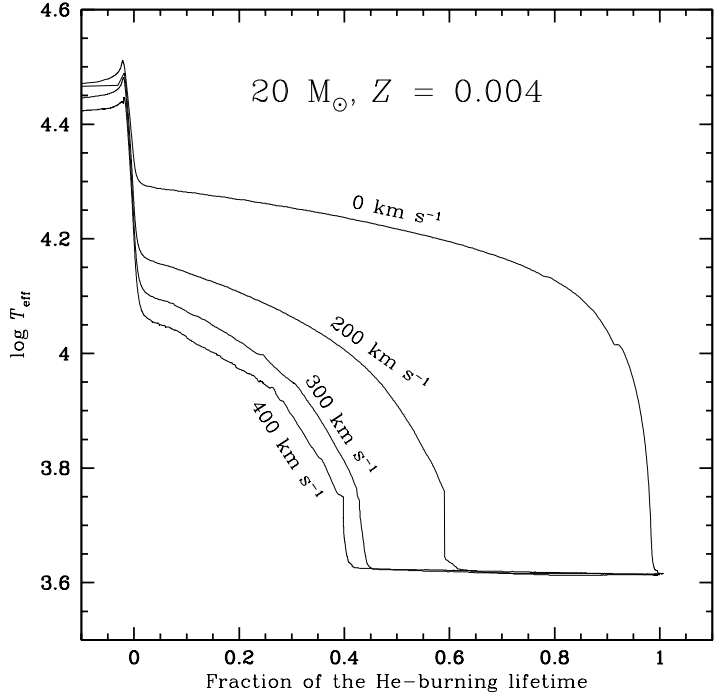}}
  \caption{Evolution of the $T_{\mathrm{eff}}$
as a function of the fraction of the lifetime spent
in the He--burning phase for 20 M$_\odot$ stars with different
initial velocities. 
}
  \label{rsgSMC}
\end{center}  
\end{figure}

Rotation, by changing the chemical structure of the star, modifies
its evolution. For instance, moderate rotation at metallicities of the Small Magellanic Cloud (SMC)
favors redward evolution in the Hertzsprung-Russel diagram. This behavior is illustrated in Fig.~\ref{rsgSMC} and can account for the high number of red supergiants observed in the SMC (\cite{MMVII}),
an observational fact which is not at all reproduced by non-rotating stellar models.

Now it is well known that the mass loss rates are greater
when the star evolves into the red part of the HR diagram, thus in this case, rotation modifies 
the mass loss indirectly, by changing the evolutionary tracks. 
The $\upsilon_{\rm ini}=0$, 200, 300 and 400 km s$^{-1}$ models lose respectively 0.14, 1.40, 1.71 and 1.93 M$_\odot$ during the core He-burning phase (see Table~1 in \cite{MMVII}). The enhancement of the mass lost reflects the longer lifetimes of the red supergiant phase when velocity increases.
Note that these number were obtained assuming that the same scaling law between mass loss and metallicity
applies during the red supergiant phase. If, during this phase, mass loss comes from continuum-opacity driven
wind then the mass-loss rate will not depend on metallicity (see the review by \cite{vL06}).
In that case, the redward evolution favored by rotation would have a greater impact on mass loss than
that shown by the computations shown above.

Of course, such a trend cannot continue forever. For instance, 
at very high rotation, the star will have a homogeneous evolution and will never become a red supergiant
(\cite{M87}).
In this case, the mass loss will be reduced, although this effect will be somewhat  compensated by two
other processes: first by the fact that the Main-Sequence lifetime will last longer and, second,
by the fact that the star will enter the Wolf-Rayet phase (a phase with high mass loss rates) at an earlier stage of its evolution.

% fig. 1

\section{Radiation driven stellar winds with rotation}

The effects of rotation on the radiation driven stellar winds 
result from the changes brought by rotation to the stellar surface. They induce changes of the morphologies
of the stellar winds and increase their intensities.

\subsection{Stellar wind anisotropies}

\begin{figure}[tb]
\begin{center}
  \resizebox{7cm}{!}{\includegraphics{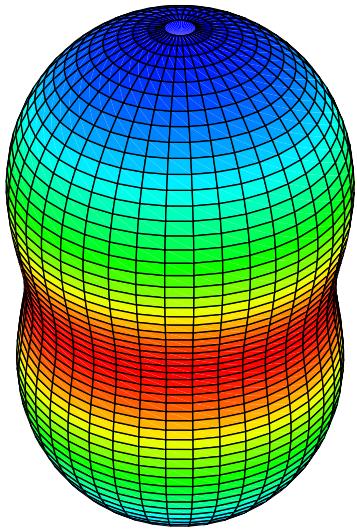}}
  \caption{Iso-mass loss distribution for a 120 M$_\odot$ star with Log L/L$_\odot$=6.0 and T$_{\rm eff}$ = 30000 K rotating at a fraction 0.8 of critical velocity. 
}
  \label{ani}
\end{center}  
\end{figure}

Naively we would first guess that a rotating
      star would lose mass preferentially from the equator, where the effective gravity (gravity decreased
      by the effect of the centrifugal force) is lower.
      This is probably true when the star reaches the critical limit (i.e. when the equatorial surface
      velocity is such that the centrifugal acceleration exactly compensates the gravity), but this is not
      correct when the star is not at the critical limit. Indeed as recalled above, a rotating star has a
      non uniform surface brightness, and the polar regions are those which have the most powerful radiative 
      flux. Thus one expects that the star will lose mass preferentially along the rotational axis. This is
      correct for hot stars, for which the dominant source of opacity is electron scattering. In that
      case the opacity only depends on the mass fraction of hydrogen and does not depends on other
      physical quantities such as temperature. Thus rotation induces 
      anisotropies of the winds   (\cite{MD01};\cite{DO02}).
      This is illustrated in Fig.~\ref{ani}.
      Wind anisotropies have consequences for the angular momentum that a star retains in its interior.
      Indeed, when mass is lost preferentially along the polar axis, little angular momentum is lost.
      This process allows loss of mass without too much loss of angular momentum a process which might
      be important in the context of the evolutionary scenarios leading to Gamma Ray Bursts. Indeed 
      in the framework of the collapsar scenario (\cite{W93}), 
      one has to accommodate two contradictory requirements: on one side, the progenitor needs to lose mass
      in order to have its H and He-rich envelope removed at the time of its explosion, and on the other hand
      it must have retained sufficient angular momentum in its central region to give birth to a fast
      rotating black-hole.
      
% fig. 2          
      
\subsection{Intensities of the stellar winds}      
      
      The quantity of mass lost through radiatively driven stellar winds is enhanced by rotation. This enhancement can occur through two channels: by reducing the effective gravity at the surface of the star, by
      increasing the opacity of the outer layers through surface metallicity enhancements due to rotational mixing.
      
\begin{itemize}

\item{\it reduction of the effective gravity: } The ratio of the mass loss rate of a star with a surface angular velocity $\Omega$ to that
      of a non-rotating star, of the same initial mass, metallicity and lying at the same position in the
      HR diagram is given by (\cite{MMVI})
      
      \begin{equation}
\frac{\dot{M} (\Omega)} {\dot{M} (0)} \simeq
\frac{\left( 1  -\Gamma\right)
^{\frac{1}{\alpha} - 1}}
{\left[ 1 - 
\frac{4}{9} (\frac{v}{v_{\mathrm{crit, 1}}})^2-\Gamma \right]
^{\frac{1}{\alpha} - 1}} \; ,
\end{equation}
\noindent
where $\Gamma$ is the electron scattering opacity for a non--rotating
star with the same mass and luminosity, $\alpha$ is a force multiplier (\cite{La95}). 
The enhancement factor remains modest for stars with luminosity sufficiently far away from the
      Eddington limit (\cite{MMVI}). Typically, $\frac{\dot{M} (\Omega)} {\dot{M} (0)} \simeq 1.5$
      for main-sequence B--stars.
      In that case, when the surface velocity approaches the critical limit, the effective
      gravity decreases and the radiative flux also decreases. Thus the matter becomes less bound
      when, at the same time, the radiative forces become also weaker. 
      When the stellar luminosity approaches the Eddington limit, the mass loss increases can be much greater,
      reaching orders of magnitude.
This comes from the fact that rotation lowers the maximum luminosity or the Eddington luminosity of a star.  
      Thus it may happen that for a velocity still 
      far from the classical critical limit, the 
      rotationally decreased maximum luminosity becomes equal 
      to the actual luminosity of the star. 
      In that case, strong mass loss ensues and the star is said to have reached
      the $\Omega\Gamma$ limit (\cite{MMVI}).

\item {\it Effects due to rotational mixing: }
During the core helium burning phase, at low metallicity,
the surface may be strongly enriched in both H-burning and He-burning products, {\it i.e.} mainly in nitrogen, carbon and oxygen. Nitrogen is produced by transformation of the carbon and oxygen produced in the He-burning core and which have diffused by rotational mixing in the H-burning shell (\cite{MMVIII}). Part of the carbon and oxygen produced in the He-core also diffuses up to the surface. Thus at the surface, one obtains very high value of the CNO elements. For instance a 60 M$_\odot$ with Z=$10^{-8}$ and $\upsilon_{\rm ini}=800$ km s$^{-1}$ has, at the end of its evolution, a CNO content at the surface equivalent to 1 million times its initial metallicity! In the present models, we have applied the usual scaling laws linking the surface metallicity
to the mass loss rates (see Eq.~1). In that case, one obtains that the star loses due to this process
more than half of its initial mass (see Table 1).

\end{itemize}
      
\section{Mechanical winds induced by rotation}      
      
As recalled above, during the Main-Sequence phase the core contracts
      and the envelope expands. In case of local conservation of the angular momentum, the core would thus
      spin faster and faster while the envelope would slow down. In that case, it can be easily shown that the surface velocity would evolve away from the critical velocity (see e.g. \cite{Vl06}). 
      In models with shellular rotation however
      an important coupling between the core and the envelope is established through the action of the
      meridional currents. As a net result, angular momentum is brought from the inner regions to the outer ones. Thus, would the star lose no mass by radiation driven stellar winds (as is the case at low Z), one expects that the surface velocity
      would increase with time and would approach the critical limit (see Fig.~3). In contrast, 
      when radiation driven stellar winds are important, the timescale for removing mass 
      and angular momentum at the surface
      is shorter than the timescale for accelerating the outer layers by the above process and the surface
      velocity decreases as a function of time. It evolves away from the critical limit. 
      Thus, an interesting situation occurs: when the star loses
      little mass by radiation driven stellar winds, it has more chance to lose mass by a mechanical wind. On the other hand, when the star loses mass at a high rate by
      radiation driven mass loss then it has no chance to reach the critical limit and thus to undergo a 
      mechanical wind. We discuss further below the possible importance of this mechanical wind.
      
\begin{figure}[tb]
\begin{center}
  \resizebox{10cm}{!}{\includegraphics{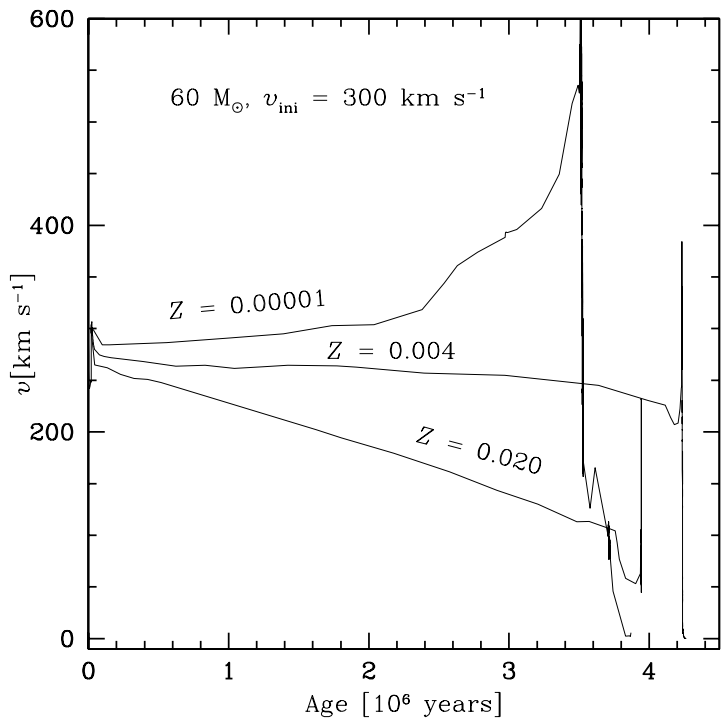}}
  \caption{Evolution of the surface velocities for a 60 M$_
 {\odot}$ star with 3 different
 initial metallicities. 
}
  \label{V60Z}
\end{center}  
\end{figure}% fig. 3      
      
\section{Discussion}  

At this point it is interesting to discuss three aspects of the various effects described above. First what
are the main uncertainties affecting them?  Second, what are their relative importance? And finally what are their consequences for the interstellar medium enrichment?

\subsection{Uncertainties}

In addition to the usual uncertainties affecting the radiation driven mass loss rates, the above processes 
poses three additional problems:
\begin{enumerate}

\item {\it What does happen when the CNO content of the surface increases by six orders of magnitude as was obtained
in the 60 M$_\odot$ model described above?} Can we apply the usual scaling law between Z and the mass losses?
This is what we have done in our models (using $\alpha=0.5$), but of course this should be studied in more details by stellar winds models. For instance, for WR stars, \cite{V05} have shown that at $Z=Z_\odot/30$, 60\% of the driving is due to CNO elements and only 10\% to Fe.
Here the high CNO surface enhancements result
from rotational mixing which enrich the radiative outer region of the star in these elements, but also from the fact that the star evolves to the red part of the HR diagram, making an outer convective zone to appear. This
convective zone plays an essential role in dredging up the CNO elements at the surface. Thus what is needed
here is the effects on the stellar winds of CNO enhancements in a somewhat red part of the HR diagram
(typical effective temperatures of the order of Log T$_{\rm eff}\sim$3.8).

\item {\it Do stars can reach the critical limit?} For instance, \cite{BO70}
obtain that during pre-main sequence evolution of rapidly rotating massive stars, ``equatorial mass loss'' or ``rotational mass ejection'' never occur (see also \cite{BO73}). In these models the condition of zero effective gravity is never reached.  However, these authors studied pre-main sequence evolution and made different hypotheses on the
transport mechanisms than in the present work. Since they were interested in the radiative contraction
phase, they correctly supposed that ``the various instabilities and currents which transport angular momentum
have characteristic times much longer than the radiative-contraction time''. This is no longer the case 
for the Main-Sequence phase. In our models, we consistently accounted for the transport of the angular
momentum by the meridional currents and the shear instabilities.
A detailed account of the transport mechanisms shows that they are never able to prevent the star from reaching
the critical velocity.
Another difference between the approach in the work of \cite{BO70} and ours is that
\cite{BO70} consider another distribution of the angular velocity than in our models. They supposed constant $\Omega$
on cylindrical surface, while here we adopted, as imposed by the theory of \cite{Z92}, a ``shellular rotation law''. They resolved the Poisson equation for the gravitational potential, while here we adopted the Roche model. Let us note that the Roche approximation appears justified in the present case, since only the outer layers, containing little mass, are approaching the critical limit. The majority of the stellar mass has a rotation rate much below the critical limit and is thus not strongly deformed by rotation. Thus these differences probably explain why in our models we reach situations where the effective gravity becomes zero.

\item {\it What does happen when the surface velocity reaches the critical limit?} 
Let us first note that when the surface reaches the critical velocity, the energy which is still needed to
make equatorial matter to escape from the potential well of the star is still important. This is because the gravity of the system continues of course to be effective all along the path from the surface to the infinity
and needs to be overcome.
If one
estimates the escape velocity from the usual equation energy for a piece of material of mass $m$ at the equator
of a body of mass $M$, radius $R$ and rotating at the critical velocity,
\begin{equation}
{1 \over 2}m \upsilon_{\rm crit}^2+{1 \over 2}m \upsilon_{\rm esc}^2-{GMm \over R}=0,
\end{equation}
one obtains, using $\upsilon_{\rm crit}^2={GM/R}$  that the escape velocity is 
simply reduced by a factor $1/\sqrt{2}=0.71$ with respect to the escape velocity from a non-rotating body
\footnote{We suppose here that
the vector $\upsilon_{\rm esc}$ is normal to the direction of the vector $\upsilon_{\rm crit}$.}.
Thus the reduction is rather limited and one can wonder if matter will be really lost.
A way to overcome this difficulty is to consider the fact that, at the critical limit, the matter will
be launched into a keplerian orbit around the star.
Thus, probably, when the star reaches
the critical limit an equatorial disk is formed like for instance around Be stars. 
Here we suppose that this disk will eventually dissipate by radiative effects and thus that the material will be lost by the star.

Practically, in the present models, we remove the supercritical layers. This removal of material allows the outer layers to become again subcritical at least until secular evolution will bring again the surface near the critical limit (see \cite{MEM06} for more details
in this process). Secular evolution during the Main-Sequence phase triggers two counteracting effects: on one side, the stellar surface expands. Local conservation of the angular momentum makes the surface to slow down and the surface velocity to evolve away from the critical limit. On the other
hand, meridional circulation continuously brings angular momentum to the surface and accelerates the outer layers. This last effect in general overcomes the first one and the star rapidly reach again the critical limit. How much mass is lost by this process?
As seen above, the two above processes will maintain the star near the critical limit for most of the time.
In the models, we adopt the mass loss rate required 
to maintain the star at about 95-98\% of the critical limit. Such a mass loss rate is imposed  as long as the secular evolution brings back the star near the critical limit. In general, during the
Main-Sequence phase, once the critical limit is reached, the star remains near this limit for the rest
of the Main-Sequence phase. At the end of the Main-Sequence phase, evolution speeds up and
the local conservation of the angular momentum overcomes the effects due to meridional currents, the star
evolves away from the critical limit and the imposed ``critical'' mass loss is turned off.

%One can wonder also, as was indicated by REF, if a convective zone could not
%appear in the equatorial region when the star is near the critical limit. If true, the presence of such a
%convective may have important impact on the mass loss triggered by this process
\end{enumerate}

\subsection{Importance of the various effects on mass loss induced by rotation}

The processes which are the most important for metal poor stars are the reaching of the critical limit
(both the classical limit and the $\Omega\Gamma$-limit) and the increase of the surface metallicity by
the concomitant effect of rotational diffusion and dredging-up by an outer convective zone.

In order to quantify the importance of the various effects discussed above, we compare in Table~1 four
60 M$_\odot$ with an initial velocity of 800 km s$^{-1}$ at four different metallicities, $Z=0$
(\cite{EMM06}), 10$^{-8}$,
10$^{-5}$ (\cite{MEM06}) and $10^{-3}$ (Decressin et al., submitted) and we give the mass lost during the MS and the post MS (PMS) phases. The mass lost by non-rotating models is also given.

\begin{table}
\begin{center}
\begin{tabular}{|l|l|l|l|}
\hline
           &                            &                       &                         \\
Z          & $\Omega/\Omega_{\rm crit}$ & $\Delta$ M$_{\rm MS}$ & $\Delta$ M$_{\rm PMS}$  \\
           &                            &                       &                         \\
\hline 
0          & 0                          & 0                     & 0.0013 (0)              \\
0          & 0.71                       & 2.42                  & 0.27   (0)              \\
           &                            &                       &                         \\
10$^{-8}$  & 0                          & 0.18                  & 0.09   (0)              \\
10$^{-8}$  & 0.77                       & 2.38                  & 33.80  (0.85)           \\
           &                            &                       &                         \\
10$^{-5}$  & 0                          & 0.21                  & 0.22   (0)              \\
10$^{-5}$  & 0.90                       & 6.15                  & 16.94  (0)              \\
           &                            &                       &                         \\                   0.0005     & 0                          & 0.78                  & 13.29  (0)              \\
0.0005     & 0.94                       & 20.96                 & 21.79  (17.15)          \\
           &                            &                       &                         \\            
\hline
\end{tabular}
\caption{Mass lost in solar masses by 60 M$_\odot$ non-rotating and rotating models at different
metallicities during the MS and the post MS phases. The number in parenthesis in the last column
indicates the mass lost during the WR phase. See text for the references of the stellar models.} 
\end{center}
\end{table}

From Table~1, we first note that a given value of the initial velocity (here 800 km s$^{-1}$)
corresponds to lower value of $\Omega/\Omega_{\rm crit}$ at lower metallicity. This is a consequence
of the fact that stars are more compact at low $Z$. Would we have kept $\Omega/\Omega_{\rm crit}$
constant one would have higher values of $\upsilon_{\rm ini}$ at low Z. 

During the MS phase, we see that the non-rotating models lose nearly no mass. 
The rotating models, on the other hands, lose some mass when reaching the critical limit.
For the Pop III star the critical limit
is reached when the mass fraction of hydrogen at the center, $X_c$, is 0.35. For the models at Z= $10^{-8}$,
$10^{-5}$ and 0.0005, the critical limit is reached respectively when $X_c$ is equal to 0.40, 0.56 and
0.65. Thus at higher metallicity, the critical limit is reached earlier.
This behavior comes from two facts: first keeping $\upsilon_{\rm ini}$ constant implies higher
$\Omega/\Omega_{\rm crit}$ at higher $Z$, then, meridional currents, which accelerate the
outer layers are more rapid at higher metallicities.

The mass lost after the Main-Sequence phase remains very modest for non-rotating stars, except for the model
at $Z=0.0005$. For the rotating models, except in the case of the Pop III models, all models lose great
amounts of material. In the case of the models with $Z=10^{-8}$ and $10^{-5}$, the main effect responsible for
the huge mass loss is the surface enrichments in CNO elements. In the case of the $Z=0.0005$, no such effect is observed, however the star, as a result of the high mass loss during the MS phase and also
due to rotational mixing, has a long WR phase, during which most of the mass is lost. The Pop III model
on the other hand loses little amount of mass during the post-MS phase. This comes from the fact that
the star evolves only at the very end of its evolution in the red part of the HR diagram, preventing thus an efficient dredging up of the CNO elements at the surface. Thus the surface enhancements remain modest and occur during a too short phase for having an important impact on mass loss. On the other hand, it would be interesting to compute models with higher initial values of  $\Omega/\Omega_{\rm crit}$.

As a general conclusion, we see that the quantity of mass lost very much depends on rotation in metal poor regions.
Moreover, the lost material is enriched in new synthesized elements like helium, carbon, nitrogen and oxygen and thus will participate to the chemical evolution of the interstellar medium. 
Short comments on this point are made in the paragraph below.

\subsection{Interesting consequences}

The effects of rotation in metal poor stellar models have an impact on the stellar populations and the nucleosynthesis. Some of these effects are mainly due to the more efficient mixing obtained in metal poor stars and do not much depend on the mass loss induced by rotation, others are consequences of both effects.

Among the effects mainly due to enhanced rotational mixing, let us mention the fact that fast rotating massive stars might be very efficient sources of primary nitrogen in metal poor regions (\cite{CMB05}; \cite{CHM06} and
see the contribution by Chiappini et al. in this volume) and lead to different trends for C/O and N/O at very
low metallicity. 
Other isotopes such as $^{13}$C, $^{18}$O could also be produced abundantly in such models.

Interesting consequences resulting from both enhanced rotational mixing and mass loss are:
\begin{itemize}
\item the possibility to explain the origin of the peculiar abundance pattern exhibited by the extremely metal poor
C-rich stars. These stars could be formed from wind material of rotating massive stars or from material ejected, either in a mass transfer episode or by winds, from a rotating E-AGB star (\cite{MEM06}, Hirschi, submitted).

\item to provide an explanation for the origin of the helium-rich stars in $\omega$ Cen. The presence of a blue
ZAMS sequence in this cluster  (in addition to a red sequence which is about a factor two less rich in iron) is interpreted as the existence in this cluster of very helium rich stars. Typically stars on the blue sequence would have, according to stellar models, a mass fraction of helium of 0.40, while stars on the red
sequence would only have an helium mass fraction of 0.25 (\cite{No04}). We have proposed that the helium-rich stars could be formed from wind material of fast rotating massive stars (\cite{MMom}). This material would indeed have the appropriate chemical composition for accounting for the abundance patterns observed in the blue sequence.

\item Interestingly, fast rotating massive stars, losing mass at the critical limit, could also contribute
in providing material for forming second generation stars  in globular clusters. Such stars
would present peculiar surface abundances, relics of their nuclear-processing in the fast
rotating massive stars (see \cite{D06}, and also the contributions by Charbonnel and Prantzos in the present volume).

\end{itemize}

\section{Conclusions}

It is now a well known fact that rotation is a key feature of the evolution of stars. For metallicities
between those of the Small Magellanic Cloud and of the solar neighborhood, rotating models better reproduce the observed characteristics of stars than non-rotating models (see e.g. the discussion in  \cite{MEM06}). From this argument alone, one would expect that
they would do the same in metal poor regions. Of course at very low metallicity, direct comparisons between
massive star models and observed stars are no longer possible. Thus it is particularly important in this case
to first check the models in metallicity range where such comparisons are possible. 
When the same physical processes as those necessary to obtain good fits at high metallicity are accounted for in metal poor stars, one notes that stars are on one hand strongly mixed and on the other hand may lose large amounts of material. These features might be very helpful for explaining numerous puzzling observational facts
concerning metal poor stars as seen above. Moreover, such models stimulate new questions which can
be the subjects of future works:
 
\begin{itemize}
\item One can wonder what would be the contribution of very fast rotating Pop III massive stars to the ionizing flux. These stars would follow an homogeneous evolution, 
evolve in the blue side of the HR diagram, would have their lifetimes
increased and would become WR stars. For all these reasons, they would likely be important sources of ionizing photons.

\item What would be the ultimate fate of such fast rotating Pop III stars? Would they give birth to collapsars as proposed by \cite{YL05} and \cite{WH06}? 

\item If pair instability supernovae have left no nucleosynthetic sign of their existence in observed metal
poor star, is this
because no sufficiently high massive stars have ever formed? Or, if formed, might these stars have skipped the
Pair Instability regime due to strong mass loss triggered by fast rotation?

\item Does the dynamo mechanism of Tayler-Spruit work in Pop III stars? This mechanism needs a small
pristine magnetic field which will be amplified at the expense of the differential rotational energy.
But does this pristine magnetic field exist in this case? 
In case the mechanism is working, how its effects vary as a function of the metallicity?

\item Could the first generations of massive stars be important producers of helium as was suspected
long time ago by \cite{TH64}? 

\item What was the distribution of the rotational velocities at different metallicities? 
Is this distribution the same in the field and in dense clusters (like globular clusters)?
As recalled above, some indirect observations indicate that the distribution might be biased toward
fast rotators at low metallicity.  
Are there any other indirect hints supporting this view? What would be the
physical mechanism responsible for such a trend? (Shorter disk locking episode in metal poor regions?).

\end{itemize}

The list above is not exhaustive. It simply reflects the richness of the subject which will certainly become a very fruitful area of research in the coming years.

\end{document}